\documentclass[12pt, epsf]{article}

\usepackage{latexsym}
\usepackage{amsmath}
\usepackage{amssymb}
\usepackage{amsfonts}
\usepackage{graphicx}
\usepackage{epsfig}
\usepackage{array}
\usepackage{flafter}

\newcommand{\Ex}[1]{Example~\ref{ex#1}}

\newtheorem{ExampleDef}{Example}[section]

\newcommand{\Example}[3]{
  \begin{list}{}{
      \setlength{\leftmargin}{1em}}     
    \item                               
    \small                              
    \begin{ExampleDef} \rm              
      {\bf \hspace{-1ex}: #1}           
      #2                                
      \hfill {\large \boldmath $\Box$}  
      \label{ex#3}                      
    \end{ExampleDef}
  \end{list}}

\begin{document}
\rightline{\small NSL-050901. September, 2005}
\vspace{1.em}
\begin{center}
{\Large \bf Higher-Order Nonlinear Contraction Analysis \par}
\vspace{1.5em}
{\large Winfried Lohmiller and Jean-Jacques E. Slotine \par}
{Nonlinear Systems Laboratory \\
Massachusetts Institute of Technology \\
Cambridge, Massachusetts, 02139, USA\\
{\sl wslohmil@mit.edu, jjs@mit.edu} \par}
\vspace{3em}
\end{center}

\begin{abstract}

Nonlinear contraction theory is a comparatively recent dynamic control
system design tool based on an exact differential analysis of
convergence, in essence converting a nonlinear stability problem into
a linear time-varying stability problem. Contraction analysis relies
on finding a suitable {\it metric} to study a generally nonlinear and
time-varying system.  This paper shows that the computation of the
metric may be largely simplified or indeed avoided altogether by
extending the exact differential analysis to the higher-order dynamics
of the nonlinear system. Simple applications in economics, classical
mechanics, and process control are described.

\end{abstract}


\section{Introduction}

Nonlinear contraction theory is a comparatively recent dynamic control
system design tool based on an exact differential analysis of
convergence~\cite{Lohm1}. In essence, it allows one to convert a
nonlinear stability problem into a linear time-varying stability
problem.  Contraction analysis relies on finding a suitable {\it
metric} to study a generally nonlinear and time-varying system.
Depending on the application, the metric may be trivial (identity or
rescaling of states), or obtained from physics, combination of
contracting subsystems~\cite{Lohm1}, semi-definite
programming~\cite{Lohm4}, or recently sums-of-squares
programming~\cite{Parrilo}. 

The goal of this paper is to show that the computation of the metric
may be largely simplified or avoided altogether by extending the exact
differential analysis to the higher-order dynamics of the nonlinear
system.  Intuitively this is not surprising, since, as an elementary
instance, a scalar linear time-invariant system would require in the
original approach a non-identity metric (obtained from a Lyapunov
Matrix Equation).

After a brief review of contraction theory in Section 2, the main
results are presented in Section 3, first in the discrete-time case
(with a simple application to price dynamics in economics) and then in
the continuous-time case. Simple examples and applications are
discussed in Section 4, in the contexts of classical mechanics,
process control, and observer design (see
\cite{Rouchon,Nguyen,Jouff,inertial} for other recent applications of
contraction theory to observer design). Hamiltonian systems are
studied in section 5. Concluding remarks are offered in section 6.

\section{Contraction theory}

The basic theorem of contraction analysis~\cite{Lohm1} can be
stated as

\newtheorem{theorem}{Theorem}
\begin{theorem} Consider the deterministic system $ \ \dot{\bf x} =
{\bf f}({\bf x},t) \ $, where ${\bf f}$ is a smooth nonlinear
function. If there exist a uniformly positive definite metric

${\bf M(\bf x}, t) \ = \ {\bf \Theta}({\bf x}, t)^T \ {\bf
\Theta}^{\ast}({\bf x}, t)$

\noindent such that the Hermitian part of the associated generalized
Jacobian
${\bf F} \ = \ \left(\dot{\bf \Theta} + {\bf \Theta}
\frac{\partial {\bf
  f}} {\partial {\bf \bf x}} \right){\bf \Theta}^{-\ast}$

\noindent is uniformly negative definite, then all system
trajectories converge exponentially to a single trajectory,
with convergence rate $| \lambda_{max} |$, where $\lambda_{max}$
is the largest eigenvalue of the Hermitian part of $\ {\bf F}$. The
system is said to be contracting. \label{th:theoremF}
\end{theorem}

In the above, $\ ^{\ast}$ denotes complex conjugation, $\ ^{-\ast}$
for a matrix denotes the inverse of the conjugate matrix, and the
state-space is $R^n$ (in this paper) or $C^n$. The system is said to
be {\it semi-contracting} (for the metric ${\bf M(\bf x}, t)$) if
${\bf F}$ is always negative semi-definite, and {\it indifferent} if
${\bf F}$ is always zero.

It can be shown conversely that the existence of a uniformly positive
definite metric with respect to which the system is contracting is
also a necessary condition for global exponential convergence of
trajectories.  In the linear time-invariant case, a system is globally
contracting if and only if it is strictly stable, with ${\bf F}$
simply being a normal Jordan form of the system and ${\bf \Theta}$ the
coordinate transformation to that form. Conceptually, approaches
closely related to contraction, although not based on differential
analysis, can be traced back to \cite{Hart} and even to \cite{Lew}.

Similarly, a discrete system 
$$ {\bf x}_{i+1} = {\bf f}_i({\bf x}_i, i)
$$ 
will be contracting in a metric ${\bf \Theta}^T{\bf \Theta}^{\ast}$ if the
largest singular value of the discrete Jacobian ${\bf \Theta}_{i+1}
\frac{\partial {\bf f}_i}{\partial {\bf x}_i} {\bf \Theta}_i^{-\ast}$ is
strictly smaller than 1.  In the particular case of real autonomous
systems with identity metric, the basic contraction theorem
corresponds in the continuous-time case to Krasovkii's theorem~\cite{Sl91}, and
in the discrete-time case to the contraction mapping theorem~\cite{Bert}.

Contraction theory proofs make extensive use of {\it virtual
displacements}, which are differential displacements at fixed time
borrowed from mathematical physics and optimization theory. Formally,
if we view the position of the system at time $t$ as a smooth function
of the initial condition ${\bf x}_o$ and of time, $\ {\bf x} = {\bf
x}({\bf x}_o ,t)\ $, then $\ \delta {\bf x} = \frac{\partial {\bf
x}}{\partial {\bf x}_o} \ d {{\bf x}_o}\ $. For instance~\cite{Lohm1},
for the system of Theorem 1, one easily computes
\begin{equation}\label{differential}
\frac{d}{dt} ({\bf \Theta}  \delta {\bf x}) \ = \ {\bf F} ({\bf \Theta}  \delta {\bf x})
\end{equation}

An appropriate metric to show that the system is contracting may be
obtained from physics, combination of contracting
subsystems~\cite{Lohm1}, semi-definite programming~\cite{Lohm4}, or
sums-of-squares programming~\cite{Parrilo}. The goal of this paper is
to show that the computation of the metric may be largely simplified
or avoided altogether by considering the system's {\it higher-order}
virtual dynamics (rather than merely its first-order virtual dynamics,
as in equation (\ref{differential})).

\section{Higher-order contraction}

\subsection{The discrete-time case}

Technically, the extension to higher-order contraction is simplest in
the discrete-time case, which we discuss first. The main idea is to
construct an exponential bound on the virtual displacement $\delta
{\bf x}$ over $n$ time-steps, rather than over a single time-step as
in \cite{Lohm1}.

Consider for $i \ge 0$ the $n$-dimensional ($n \ge 1$) virtual
dynamics
$$ 
 \delta {\bf x}_{i+n} = {\bf A}^{n-1}_i \delta {\bf x}_{i+n-1} +
 \ldots + {\bf A}^o_i \delta {\bf x}_i
$$ 
Taking the norm (denoted by $|\ \ |$\ ) on both sides, and bounding, yields 
\begin{equation} 
|\delta {\bf x}_{i+n}| \ \le \ \ |A^{n-1}_i| \ |\delta {\bf x}_{i+n-1}|\ + \
\ldots \ + \ |A^o_i| \ |\delta {\bf x}_i| \nonumber
\end{equation}
where the norm of a matrix is the largest singular value of that
matrix. Let us bound for $i=0$ the initial conditions using real
positive constants $\lambda$ and $K$ as
$$
  |\delta {\bf x}_{j}| \le K \lambda^{j}, \ 0 \le j < n 
$$
Assume now that the following characteristic equation is verified,
$$  
\lambda^n \ge |A^{n-1}_i| \lambda^{n-1} + \ldots + |A^o_i|, \ \forall
i \ge 0 \nonumber
$$
We then get
$$ 
|\delta {\bf x}_{i+n}| \le |A^{n-1}_i| K \lambda^{i+n-1} + \ldots +
|A^o_i| K \lambda^i \le K \lambda^{i+n}
$$ 
\begin{figure}
\begin{picture}(200,200)(0,-100)
\put(-2,0){\vector(1,0){210}}
\put(3,-100){\vector(0,1){200}}
\put(215,0){$i$}
\put(0,105){$\delta x_i$}
\put(-10,0){$0$}
\put(25,-2){$:$}
\put(50,-2){$:$}
\put(75,-2){$:$}
\put(100,-2){$:$}
\put(125,-2){$:$}
\put(150,-2){$:$}
\put(175,-2){$:$}
\put(0,30){$\bullet$}
\put(25,-30){$\bullet$}
\put(50,40){$\bullet$}
\put(75,-35){$\bullet$}
\put(100,-15){$\bullet$}
\put(125,-28){$\bullet$}
\put(150,17){$\bullet$}
\put(175,-15){$\bullet$}
\qbezier(0,90)(100,27)(200,13)
\qbezier(0,-90)(100,-27)(200,-13)
\end{picture}
\caption{$\pm K \lambda^i$ defined by $\delta x_i$ over $i$}
\label{fig:discretecontraction}
\end{figure}
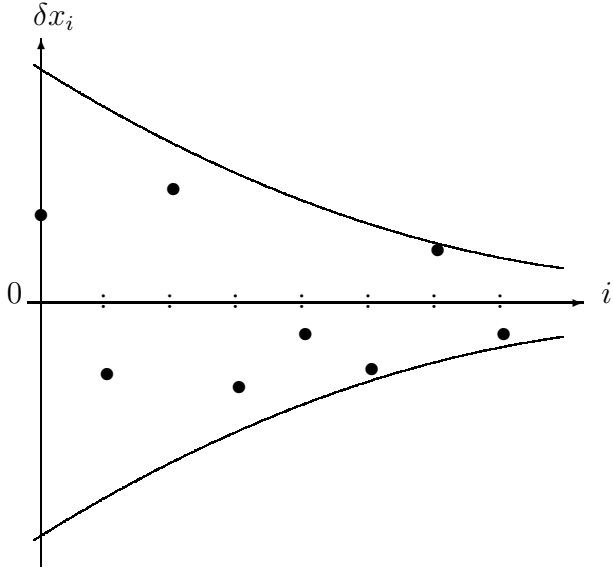
Repeating the above recursively for $ i \ge 0$ we get by complete
induction
$$ 
|\delta {\bf x}_i| \le K \lambda^i
$$
and hence exponential convergence of $|\delta {\bf x}_i|$, as
illustrated in Figure~\ref{fig:discretecontraction}.

\begin{theorem}
Consider for $i \ge 0$ the $n$-dimensional ($n \ge 1$) virtual dynamics  
$$ 
 \delta {\bf x}_{i+n}     = {\bf A}^{n-1}_i \delta {\bf x}_{i+n-1} +
 \ldots + {\bf A}^o_i \delta {\bf x}_i
$$ 
Let us define $\forall i \ge 0$ a constant $\lambda$ with the
characteristic equation
\begin{eqnarray} 
\lambda^n &\ge& |A^{n-1}_i| \lambda^{n-1} + \ldots + |A^o_i|, \
\forall i \ge 0 \ \label{eq:discretecharacteristic}
\end{eqnarray}
We can then conclude
$$
|\delta {\bf x}_{i+n}| \le K \lambda^{i+n}
$$  
where $K$ is defined by
\begin{equation}
 |\delta {\bf x}_{j}| \le K \lambda^{j}, \ 0 \le j < n \
  \label{eq:discretedefinitionofK} 
\end{equation}
Thus, the system is contracting if $\ \lambda < 1$.
\label{th:higherorderdiscrete}
\end{theorem}
\Example{}{Consider first a second-order linear time invariant (LTI)
  dynamics
$$
x_{i+2} \ + \ 2\ \gamma \ x_{i+1} \ + \ \alpha \gamma^2 x_i \ = \ u_i
$$ where $u_i$ is an input, and $\gamma$ and $\alpha$ are
constants. The virtual dynamics is
$$
\delta x_{i+2} = - \ 2\ \gamma \ \delta x_{i+1}\  - \ \alpha \gamma^2
\ \delta x_i  
$$
The characteristic equation (\ref{eq:discretecharacteristic}) for
$\lambda \ge 0$ is then given by
\begin{equation}
\lambda^2 \ \ge \ 2\ |\gamma| \ \lambda + |\alpha| \gamma^2 \ \ \ \ \
\ \ \ \ i.e. \ \ \ \ \ \ \ \ \ \lambda \ \ge \ |\gamma| (1 + \sqrt{
  1 + |\alpha |\ }) \nonumber
\end{equation}
Thus, the contraction condition $\lambda < 1$, or
$$
|\gamma| ( 1 + \sqrt{ 1 + |\alpha | \ }) \ < \ 1
$$
simply means that both eigenvalues of the system have to lie for the
conjugate complex case ($\alpha > 1$) within the red half circles in
(\ref{fig:discretecirecle}) or on the green line for the real case 
($\alpha \le 1$). Note that Theorem \ref{th:higherorderdiscrete}
simply bounds the possibly oscillating discrete system with a
non-oscillating system of the same convergence rate for the real case.

Consider now the virtual dynamics of an arbitrary second-order
nonlinear time-varying system,
$$
\delta x_{i+2} + \ 2\ \gamma(i)\ \delta x_{i+1} + \alpha(i) \
\gamma^2(i) \delta x_i \ = \ 0
$$
The characteristic equation and the contraction condition are the same as
above, except that $\ \gamma\ $ and $\ \alpha\ $ are now
time-dependent.
\begin{figure}[h]
\begin{center}
\epsfig{figure=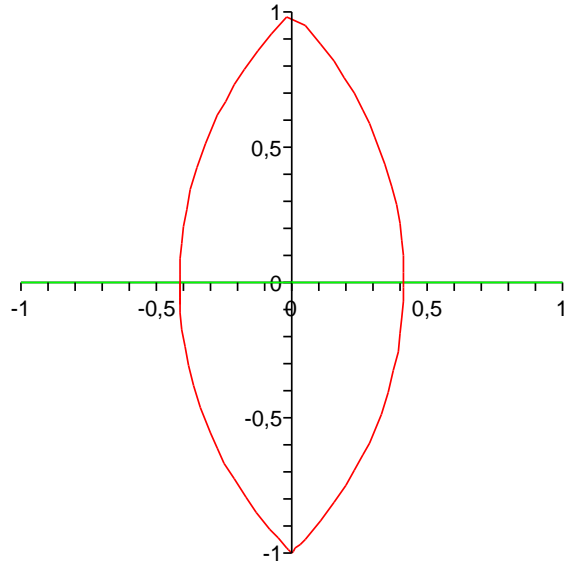,height=90mm,width=90mm}
\end{center}
\caption{Contraction region in the complex plane of second-order
  discrete system} \label{fig:discretecirecle}
\end{figure}
}{LTVdiscrete}
\Example{}{In economics, consider the price dynamics
\begin{eqnarray}
{\bf n}_{i+1} &=& {\bf f}_i({\bf p}_i, i) \nonumber \\
{\bf p}_{i+1} &=& {\bf g}_i({\bf n}_i, i) \nonumber
\end{eqnarray}
with ${\bf n}_i$ the number of sold products at time $i$ and
corresponding price ${\bf p}_i$. 

The first line above defines the customer demand as a reaction to a
given price. The second line defines the price, given by the
production cost under competition, as a reaction to the number of sold
items.  The dynamics above corresponds to the second-order economic
growth cycle dynamics
$$
{\bf n}_{i+2} = {\bf f}_{i+1} \left( {\bf g}_i ({\bf n}_i, i) \right)
$$

Contraction  behavior of this economic behavior with contraction rate
$\lambda$ can then be concluded with Theorem
\ref{th:higherorderdiscrete} for
\begin{equation}
\lambda^2 \ge | \frac{\partial {\bf f}_{i+1}}{\partial {\bf p}_{i+1}}
\frac{\partial {\bf g}_i}{\partial {\bf n}_i} | \label{eq:game}
\end{equation}
That means we get stable (contraction) behavior if the product of
customer demand sensitivity to price and production cost sensitivity
to number of sold items has singular values less than 1. We can get
unstable (diverging) behavior for the opposite case. 

Note that this result even holds when no precise model of the
sensitivity is known, which is usually the case in economic or game
situations. 

Whereas the above is well known for LTI economic models we can see
that the economic behavior is unchanged for a non-linear, time-varying
economic environment.

Let as assume now that the above corresponds to a game situation (see
e.g. \cite{Shamma} or \cite{Bryson}) between two players with
strategic action ${\bf p}_i$ and ${\bf n}_i$. Both players optimize
their reaction ${\bf g}$ and ${\bf f}$ with respect to the opponent's
action.

We can then again conclude for (\ref{eq:game}) to global contraction
behavior to a unique time-dependent trajectory (in the
autonomous case, the Nash equilibrium).}{economics}

Of course, and throughout this paper, in some cases the analysis may
yet be further streamlined by {\it first} applying a simplifying metric
transformation of the form $\ \delta {\bf z} = {\bf \Theta} \delta
{\bf x}\ $, and then applying the results to $\ \delta {\bf z}$.

\subsection{The continuous-time case}

Let us now derive the continuous-time version of the previous
results. Consider for $t\ge0$ the $n$-dimensional ($n \ge 1$) virtual dynamics  
$$ 
 \delta {\bf x}^{(n)} = - {\bf A}_{n-1} \delta {\bf x}^{(n-1)} - \
 ... \ - {\bf A}_o \delta {\bf x}
$$ 
The proof is based on splitting up the dynamics into a stable part,
described by a block diagonal matrix composed of identical negative
definite blocks ${\bf F}$ which we select, and an unstable
higher-order part. Let $\delta {\bf x}_o = \delta {\bf x}$, and define
recursively
\begin{eqnarray}
\delta \dot{\bf x}_o     &=& {\bf F} \delta {\bf x}_o + \delta {\bf
  x}_1 \nonumber \\
                        &...& \nonumber \\
\delta \dot{\bf x}_{n-2} &=& {\bf F} \delta {\bf x}_{n-2} +
\delta {\bf x}_{n-1} \nonumber \\ 
\delta \dot{\bf x}_{n-1} &=& - {\bf A}_{n-1} \delta {\bf x}_o^{(n-1)} -
{\bf A}_{n-2} \delta {\bf x}_o^{(n-2)} - \ ... \ -  {\bf A}_o \delta
{\bf x}_o \nonumber \\ 
&&- \left( {\bf F} \delta{\bf x}_o \right)^{(n-1)} - ... - \left( {\bf
    F} \delta {\bf x}_{n-2} \right)^{(1)} \nonumber \\
&=& -{\bf A}_{n-1} \left( {\bf F} \delta {\bf x}_o + \delta {\bf
  x}_1 \right)^{(n-2)} - ... - {\bf A}_o \delta {\bf x}_o \nonumber \\
&&- \left( L^1 {\bf F} \delta {\bf x}_o + L^o {\bf F} \delta {\bf x}_1
\right)^{(n-2)} - ... - \left( L^1 {\bf F} \delta {\bf x}_{n-2} + L^o
  {\bf F} \delta {\bf x}_{n-1} \right) \nonumber \\
&=& {\bf F} \delta {\bf x}_{n-1} 
- {\bf A}_{n-1}^{\ast} \delta {\bf x}_{n-1} 
- {\bf A}_{n-2}^{\ast} \delta {\bf x}_{n-2}
- {\bf A}_{n-3}^{\ast} \delta {\bf x}_{n-3} -
... \label{eq:superimposed}
\end{eqnarray}
where
\begin{eqnarray}
L^o {\bf F} &=& {\bf F} \nonumber \\
L^j {\bf F} &=& \frac{d}{dt} L^{j-1} {\bf F} + L^{j-1} {\bf F} \
{\bf F} \ \ \ \ \ \ \ \ \ j \ge 1 \nonumber
\end{eqnarray}
and
\begin{eqnarray}
{\bf A}_{n-1}^{\ast} &=& {\bf A}_{n-1} + \left( \begin{array}{c} n \\
    1 \end{array} \right) L^o {\bf F} \label{eq:Aast} \\
{\bf A}_{n-2}^{\ast} &=& {\bf A}_{n-2} + \left( \begin{array}{c} n-1
    \\ 1 \end{array} \right) {\bf A}_{n-1} L^o {\bf F} + \left(
  \begin{array}{c} n \\ 2 \end{array} \right) L^1 {\bf F} \nonumber \\
{\bf A}_{n-3}^{\ast} &=& {\bf A}_{n-3} + \left( \begin{array}{c} n-2
    \\ 1 \end{array} \right) {\bf A}^{n-2} L^o {\bf F} + \left(
  \begin{array}{c} n-1 \\ 2 \end{array} \right) {\bf A}^{n-1} L^1 {\bf
  F} + \left( \begin{array}{c} n \\ 3 \end{array} \right) L^2 {\bf F}
\nonumber \\ &...& \nonumber
\end{eqnarray}
Equation (\ref{eq:superimposed}) represents the superposition of a
higher-order-system and a block diagonal dynamics in the chosen ${\bf
F}$. Let us assess the contraction behavior of the higher-order part
by taking the norm
$$
|\delta {\bf x}^{(n)}| \le |{\bf A}_{n-1}^{\ast}| |\delta {\bf
x}^{(n-1)}| + |{\bf A}_{n-2}^{\ast}| |\delta {\bf x}^{(n-2)}| + \
\ldots \ 
$$
where the norm of a matrix is the largest singular value
of that matrix. Let us bound for $t=0$ the initial
conditions with real and constant $\lambda, K \ge 0$ and assume the
following characteristic equation
\begin{eqnarray} 
|\delta {\bf x}^{(j)}| &\le& K \lambda^j e^{\lambda t}, \ \ \ \ \ \ \ \
\ \ \ \ \ \ \ \ \ \ \ \ \ \ \ 0 \le j < n \ \label{eq:definitionofK}
\\ \lambda^n &\ge& |{\bf A}_{n-1}^{\ast}| \lambda^{n-1} + \ \ldots \ +
|{\bf A}_o^{\ast}|, \ \forall t \ge 0 \label{eq:characteristic}
\end{eqnarray}
Figure \ref{fig:discretecontraction} shows how $K$ has to be selected
for a given $\lambda$ for a second-order system ($n=2$).
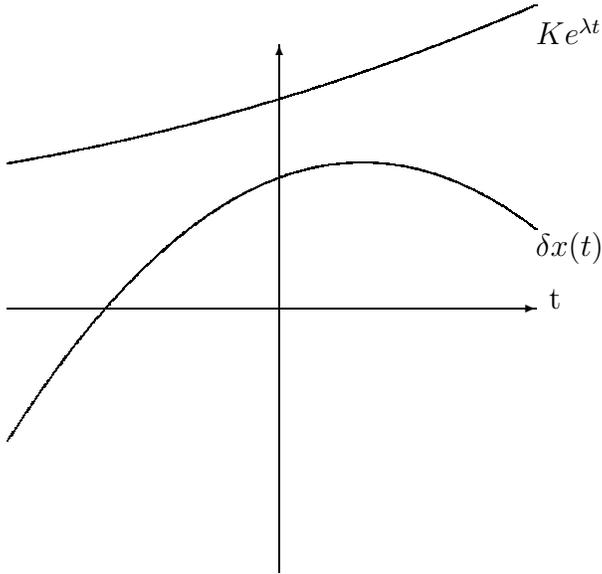
\begin{figure}
\begin{picture}(200,200)(0,-100)
\put(0,0){\vector(1,0){200}}
\put(103,-100){\vector(0,1){200}}
\put(205,0){t}
\put(200,20){$\delta x(t)$}
\put(200,100){$K e^{\lambda t}$}
\qbezier(0,-50)(100,107)(200,30)
\qbezier(0,55)(100,72)(200,115)
\end{picture}
\caption{$K e^{\lambda t}$ bounding the first and second derivative of
  $\delta x$ for a given $\lambda$ at $t=0$}
\label{fig:discretecontraction}
\end{figure}
With (\ref{eq:characteristic}) we can bound the $n$'th derivative of
$\delta {\bf x}$ as
$$
|\delta {\bf x}^{(n)}| \le |{\bf A}_{n-1}^{\ast}| K \lambda^{n-1}
e^{\lambda t} + \ \ldots \  + |{\bf A}_o^{\ast}| K e^{\lambda t} \le K
\lambda^n e^{\lambda t}
$$
Integrating the above for $t \ge 0$ we can exponentially bound the
higher-order dynamics $\delta {\bf x}$ as 
$$ 
|\delta {\bf x}| \le K e^{\int_o^t \lambda d \tau}
$$ 
Using the above this allow to conclude:
\begin{theorem}
Consider for $t\ge0$ the $n$-dimensional ($n \ge 1$) virtual dynamics  
$$ 
 \delta {\bf x}^{(n)} = - {\bf A}^{n-1} \delta {\bf x}^{(n-1)} - \
 \ldots \  - {\bf A}^o \delta {\bf x}
$$  
Let us define a constant $\lambda \ge 0$ such that $\forall t \ge 0$
we fulfill the characteristic equation
\begin{equation} 
 \lambda^n \ge {\bf A}_{n-1}^{\ast} \lambda^{n-1} + \ \ldots \  + {\bf
   A}_o^{\ast} \label{eq:theoremcharacteristic}
\end{equation}
where ${\bf A}_j^{\ast}$ is defined in (\ref{eq:Aast}) for a given choice 
of the matrix ${\bf F}$. 

We can then conclude on contraction rate (i.e., the largest eigenvalue
of the symmetric part of) ${\bf F} + \lambda {\bf I}$, where
$|\delta {\bf x}|$ is initially bounded with $K$, defined in
(\ref{eq:definitionofK}). 
\label{th:higherordercontinuous}
\end{theorem}
One specific choice of ${\bf F}$ is $-\frac{{\bf A}_{n-1}}{n}$, which
cancels the highest time-derivative on the right-hand side, and is
known for LTV systems as the reduced or unstable form~\cite{Kailath}
of the original higher-order dynamics. We will use this definition of
${\bf F}$ in most of the following examples. Also note that more
general forms could be chosen for the stable part.

\section{Examples and Applications}

In this section, we discuss simple examples (section 4.1),
applications to nonlinear observer design (section 4.2), and adding
an indifferent system (section 4.3).

\subsection{Some simple examples}

\Example{}{Consider the second-order LTI dynamics
$$
\ddot{x} = - 2 \zeta \omega \dot{x} - \omega^2 x
$$
with constant $\zeta$ and $\omega \ge 0$. The virtual dynamics is
$$
\delta \ddot{x} = - 2 \zeta \omega \delta \dot{x} - \omega^2 \delta x  
$$
The characteristic equation (\ref{eq:theoremcharacteristic}) is then
given with $F = \zeta \omega$ for constant, positive $\lambda$ by 
\begin{eqnarray}
 \lambda^2 \ge | - \omega^2 + \frac{(2 \zeta \omega)^2}{4} | =
| \zeta^2 - 1 | \omega^2 \ \ \ \ \ \ \ \ \ \rm{i.e.} \ \ \ \ \ \ \ \ \ 
\lambda \ge \omega \sqrt{| \zeta^2 - 1 |} \nonumber
\end{eqnarray}
Using Theorem \ref{th:higherordercontinuous}
we can then conclude on contraction behavior with convergence rate 
$$
\omega (-\zeta  + \sqrt{| \zeta^2 - 1 |})
$$
This means that we require the poles to lie within the $\pm45^o$ 
quadrant of the left-half complex plane.}{LTI}

While Theorem \ref{th:higherordercontinuous} can thus be
overly conservative for LTI systems, this is not the case for general
nonlinear time-varying systems, as we now illustrate.

\begin{figure}[h]
\begin{center}
\epsfig{figure=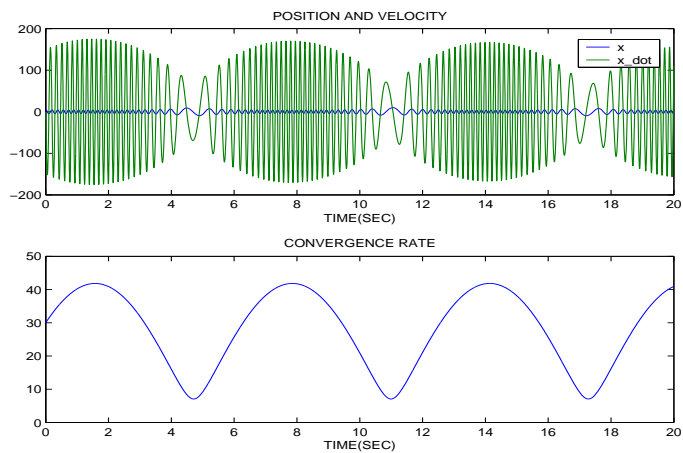,height=60mm,width=90mm}
\end{center}
\caption{\small LTV system with $a_0=900+850\sin t$ and
$a_1=0.01$} 
\label{fig:example_34}
\end{figure}

\Example{}{Consider the second-order LTV dynamics
$$
\ddot{x} + a_1 \ \dot{x} + a_o(t)\ \ x = u(t)
$$
with $a_1, a_o \ge 0$, which would be sufficient conditions for LTI
stability. Let us assume a small damping gain $a_1$ and strong spring
gains $a_o$ such that the system oscillates.

If now the time-varying gain $a_o(t)$ is chosen to be very large when
the system oscillates back to $0$ and small otherwise then the energy
is constantly increased, which makes the system unstable (Figure
\ref{fig:example_34}). This is precisely what is excluded by Theorem
\ref{th:higherordercontinuous}.}{LTV}

\begin{figure}[h]
\begin{center}
\epsfig{figure=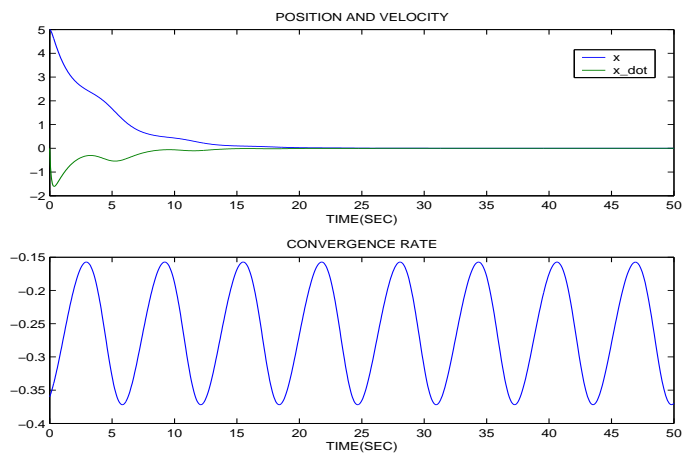,height=60mm,width=90mm}
\end{center}
\caption{\small LTV system with $a_0=2+\cos t$ and $a_1=8+\sin
t$} 
\label{fig:example_35}
\end{figure}

\Example{}{Consider the second-order LTV dynamics
$$
\ddot{x} + a_1(t)\ \dot{x} + a_o(t)\ x = u(t)
$$
The virtual dynamics is
$$
\delta \ddot{x} = - a_1 \delta \dot{x} - a_o \delta x  
$$
The characteristic equation (\ref{eq:theoremcharacteristic}) is then
given with $F = - \frac{a_1}{2}$ by 
$$
 \lambda^2 \ge | - a_o + \frac{a_1^2}{4} + \frac{\dot{a_1}}{2}|
$$ 
for constant positive $\lambda$. Using Theorem
\ref{th:higherordercontinuous} we can then conclude on contraction
behavior with convergence rate $-\frac{a_1}{2} + \lambda\ $ (Figure
\ref{fig:example_35}).}{LTV}

\begin{figure}[h]
\begin{center}
\epsfig{figure=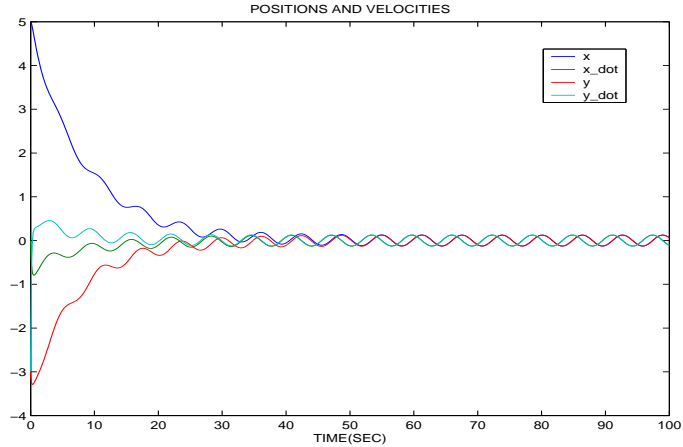,height=60mm,width=90mm}
\end{center}
\caption{\small Van der Pol with $a_1=8+\sin x$ and $u(t)=\cos
t$} 
\label{fig:example_37}
\end{figure}

\Example{}{Let us illustrate a case where choosing ${\bf F}$ other
than $\ - \frac{{\bf A}_{n-1}}{n}\ $ can simplify the result.  Consider
the generalized Van der Pol or Lienard dynamics
$$
\ddot{x} = - a_1(x) \dot{x} - a_o(x, t)
$$
with $a_1, \frac{\partial a_o}{\partial x} \ge 0$. The virtual
dynamics is
$$
\delta \ddot{x} = - a_1 \delta \dot{x} - \left( \frac{\partial
a_o}{\partial x} + \dot{a_1} \right) \delta x  
$$
The characteristic equation (\ref{eq:theoremcharacteristic}) is then
given with $F = - a_1 + \frac{\min(a_1)}{2}$ by 
$$
\lambda^2 = ( a_1 - \min(a_1) ) \lambda + |-\frac{ \min(a_1)^2}{4} +
\frac{a_1 \min( a_1)}{2} - \frac{\partial a_o}{\partial x}|   
$$ 
and hence $\lambda \ge 0.5 \left( a_1 - \min(a_1) + \sqrt{ ( a_1 -
\min(a_1) )^2 + |-\min(a_1)^2 + 2 a_1 \min(a_1) - 4 \frac{\partial
a_o}{\partial x}|} \right)$. Thus the contraction behavior of
$\delta x$ is then given by
$$
\lambda + F \le - 0.5 a_1 + 0.5 \sqrt{ ( a_1 - \min(a_1) )^2 +
|-( a_1 - \min(a_1) )^2 + a_1^2 - 4 \frac{\partial a_o}{\partial x}|} 
$$
which is negative if the argument of the square root is less than
$a_1^2$, i.e. for $\min(a_1)^2 - 2 a_1 \min(a_1) + 4 \frac{\partial
a_o}{\partial x} \le 0$. 

Hence we can conclude on contraction behavior if the poles of the
minimal damping $\min(a_1)$ with the actual spring gain
$\frac{\partial a_o}{\partial x}$ lie within the $\pm 45^o$ quadrant of
the left-half complex plane (Figure
\ref{fig:example_37}). Note that this result can be extended to
the vector case ${\bf x}$ if the corresponding matrix $A_1({\bf x})$
is integrable.}{VanderPol}

\begin{figure}[h]
\begin{center}
\epsfig{figure=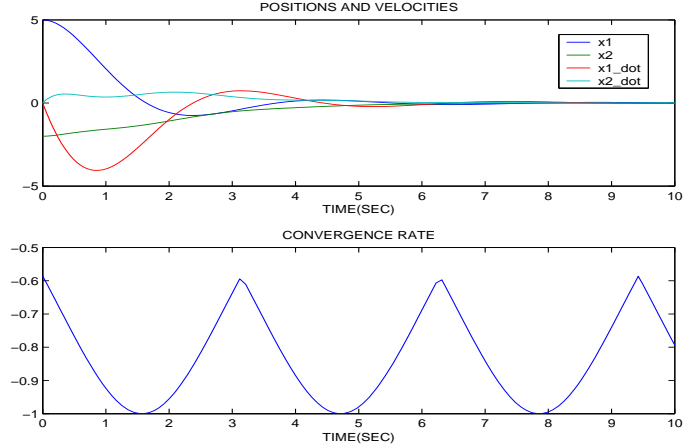,height=60mm,width=90mm}
\end{center}
\caption{\small Second-order system with $ {\bf D}=diag(1,4)$
and $V=x_1^2+x_2^2+x_1x_2\sin t$} 
\label{fig:example_36}
\end{figure}

\Example{}{Consider the second-order nonlinear vector system
$$
\ddot{\bf x} + {\bf D} \ \dot{\bf x} + \frac{\partial V}{\partial {\bf
    x}} = u(t)
$$
with potential energy $V = x_1^2+x_2^2+x_1x_2\sin t$ and constant
damping gain ${\bf D} =diag(1,4)$.

The corresponding variational dynamics is
$$
\delta \ddot{\bf x} + {\bf D} \ \delta \dot{\bf x} + \frac{\partial^2
  V}{\partial {\bf x}^2} \ \delta {\bf x} = 0
$$
The characteristic equation (\ref{eq:theoremcharacteristic}) for ${\bf
  F} = - \frac{\bf D}{2}$ is then given by 
$$
 \lambda^2 \ge | \frac{\partial^2
 V}{\partial {\bf x}^2} - \frac{\bf D D}{4}\ |
$$ 
for constant positive $\lambda$. Using Theorem
\ref{th:higherordercontinuous} we can then conclude on contraction
behavior with convergence rate $\ \lambda {\bf I}\ -\frac{\bf D}{2}$ in
Figure \ref{fig:example_36}.}{second-order-system}

\Example{}{Let us consider two
coupled systems of the same dimensions, with the virtual dynamics
$$
\frac{d}{dt}
\left(
\begin{array}{c}
\delta {\bf x}_1 \\
\delta {\bf x}_2
\end{array}
\right) \ = \
\left(
\begin{array}{cc}
{\bf F}_{11} & {\bf F}_{12}  \\
{\bf F}_{21} & {\bf F}_{22}
\end{array}
\right)
\left(
\begin{array}{c}
\delta {\bf x}_1 \\
\delta {\bf x}_2
\end{array}
\right)
$$
Let us transform this dynamics in the following second-order
dynamics
\begin{eqnarray}
\frac{d^2}{dt^2} \ \delta {\bf x}_1 &=& \dot{\bf F}_{11} \delta
{\bf x}_1 + {\bf F}_{11} \ \frac{d}{dt}\ \delta {\bf x}_1 +
(\dot{\bf F}_{12} + {\bf F}_{12} {\bf F}_{22}) \delta {\bf x}_2 +
{\bf F}_{12} {\bf F}_{21} \delta {\bf x}_1 \nonumber \\ &=& \left(
{\bf F}_{11} + {\bf F}_{22}^{\ast} \right) \frac{d}{dt} \ \delta
{\bf x}_1 + \left( \dot{\bf F}_{11} + {\bf F}_{12} {\bf F}_{21} -
{\bf F}_{22}^{\ast} {\bf F}_{11} \right) \delta {\bf x}_1 \nonumber
\end{eqnarray}
with the generalized Jacobian ${\bf F}_{22}^{\ast} = (\dot{\bf
F}_{12} + {\bf F}_{12} {\bf F}_{22}) {\bf F}_{12}^{-1}$. The
shrinking rate of this system is now the average of ${\bf F}_{11}$
and ${\bf F}_{22}^{\ast}$. Using a direct contraction approach with
e.g. ${\bf F}_{12}\ = -\ k \ {\bf F}_{21}^T$ the guaranteed
contraction rate would be a more conservative value, namely the
largest of the individual contraction rates of ${\bf F}_{11}$ and
${\bf F}_{22}^{\ast}$.}{second-order-state-space}

\subsection{Higher-order observer design}

While a controller for an $n^{\rm th}$ order system simply has to add
stabilizing feedback in ${\bf x}^{(n-1)}, \ \ldots \ , {\bf x}, t$
according to Theorem \ref{th:higherordercontinuous}, the situation is
not such straightforward for observers since here only a part of the
state is measured. Motivated by the linear Luenberger observer and the
linear reduced-order Luenberger observer, we derive such an observer
design for higher-order nonlinear systems.

Consider the $n$-dimensional nonlinear system dynamics
$$
{\bf x}^{(n)} = {\bf a}_o({\bf x}, \dot{\bf x}, t) + \ \ldots \  + {\bf
a}_{n-2}^{(n-2)} ({\bf x}, \dot{\bf x}, t) + {\bf
a}_{n-1}^{(n-1)}({\bf x}, t)
$$
with the measurement ${\bf y}({\bf x}^{(n-1)}, \ \ldots \ , {\bf x},
t)$. Note that for a linear Luenberger observer ${\bf y}$ is
equivalent to ${\bf x}$ and all ${\bf a}_i$ are linear functions of
${\bf x}$.

Consider now the corresponding nonlinear observer 
\begin{eqnarray}
\dot{\hat{\bf x}}_{n-1} &=& {\bf a}_o - {\bf e}_{o}(\hat{\bf y}) +
{\bf e}_{o}({\bf y}) \nonumber \\ \dot{\hat{\bf x}}_{n-2} &=& \hat{\bf
x}_{n-1} + {\bf a}_1 - {\bf e}_{1}(\hat{\bf y}) + {\bf e}_{1}({\bf y})
\nonumber \\ &\ \ldots \ & \nonumber \\ \dot{\hat{\bf x}}_o &=&
\hat{\bf x}_1 + {\bf a}_{n-1}- {\bf e}_{n-1}(\hat{\bf y}) + {\bf
e}_{n-1}({\bf y})\nonumber
\end{eqnarray}
with ${\bf a}_i(\hat{\bf x}_o, \hat{\bf x}_1 + {\bf a}_{n-1}- {\bf
e}_{n-1}(\hat{\bf y}) + {\bf e}_{n-1}({\bf y}), t)$ and ${\bf x}_o =
{\bf x}$.  Note that the coordinate transformation in the bracket is a
nonlinear generalization of the reduced Luenberger observer.  The
above dynamics is equivalent to
\begin{eqnarray}
\hat{\bf x}^{(n)} &=& {\bf a}_o(\hat{\bf x}, \dot{\hat{\bf x}}, t) + \
\ldots \ + {\bf a}_{n-2}^{(n-2)} (\hat{\bf x}, \dot{\hat{\bf x}}, t) +
{\bf a}_{n-1}^{(n-1)}(\hat{\bf x}, t) \nonumber \\ &-& {\bf
e}_o(\hat{\bf y}) + {\bf e}_o({\bf y}) -\ \ldots \ - {\bf
e}_{n-1}^{(n-1)}(\hat{\bf y}) + {\bf e}_{n-1}^{(n-1)}({\bf y})
\nonumber
\end{eqnarray}
whose variational dynamics is
$$
\delta \hat{\bf x}^{(n)} = \delta {\bf a}_o + \ \ldots \  + \delta
{\bf a}_{n-1}^{(n-1)} - \delta {\bf e}_o -\ \ldots \ - \delta {\bf
  e}_{n-1}^{(n-1)}
$$
where the varation is performed on $\hat{\bf x}^{n-1}, \ \ldots \ ,
\hat{\bf x}$. Hence the feedback is not only performed in $\hat{\bf
y}$, but implicitly also up to the $(n-1)^{\rm th}$ time-derivative of
$\hat{\bf y}$.

\Example{}{Consider the second-order nonlinear system
$$
\ddot{x} + \frac{\partial a_1}{\partial x} (x)\ \dot{x} + a_o(x) = 0
$$
where $x$ is measured. Consider now the corresponding nonlinear 
observer 
\begin{eqnarray}
\dot{\hat{x}}_1 &=& a_o(\hat{x})       - e_o (\hat{x} - x) 
\nonumber \\ 
\dot{\hat{x}}_o &=& \hat{x}_1 + a_1(x) - e_1 (\hat{x} - x) 
\nonumber
\end{eqnarray}
with constant $e_o$ and $e_1$ and where we have replaced with the
feedback $a_1$ as a function of $x$. The corresponding second-order 
variational dynamics is
$$
\delta \ddot{\hat{x}} + e_1 \ \delta \dot{\hat{x}} + (e_o -
\frac{\partial a_o}{\partial \hat{x}})\ \delta \hat{x} = 0
$$
Contraction behavior can then be shown with Theorem
\ref{th:higherordercontinuous}.}{second-order-system-observer} 

\begin{figure}[h]
\begin{center}
\epsfig{figure=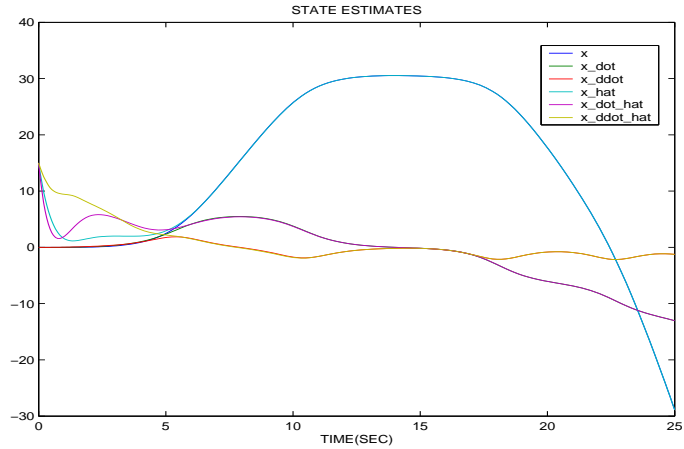,height=60mm,width=90mm}
\end{center}
\caption{\small $\dddot{x}\ = \ \sin (\dot{x})+0.1\sin
t-0.015\ $, with $\ y=x$.}\label{fig:example_4}
\end{figure}

\Example{}{Consider the temperature-dependent reaction
$A \rightarrow B$ in a closed tank 
$$
\frac{d}{dt}
\left( 
\begin{array}{c}
c_A \\
T 
\end{array}
\right) =
\left( 
\begin{array}{c}
-1 \\
-10
\end{array}
\right)
e^{-\frac{E}{T}} c_A
$$
with $c_A$ the concentration of A, $T$ the measured temperature, and
$E$ the specific activation energy. This reaction dynamics is
equivalent to the following second-order dynamics in temperature
$$
\ddot{T} + \frac{-E}{T^2} \dot{T}^2 = -e^{-\frac{E}{T}} \dot{T} 
$$
Letting $\tau = \int_o^T e^{\frac{-E}{T}} dT$ yields
$$
\ddot{\tau} = -e^{-\frac{E}{T}} \dot{\tau} 
$$
Contraction can then be shown as in \Ex{VanderPol}.
The observer dynamics
\begin{eqnarray}
\dot{\hat{T}}_1 &=& -e^{-\frac{E}{\hat{T}_o}} \dot{\hat{T}}_o +
\frac{E}{\hat{T}_o^2} \dot{\hat{T}}_o^2 - e_o e^{\frac{E}{\hat{T}_o}} 
\int^{\hat{T}}_T e^{\frac{-E}{T}} dT \nonumber \\
\dot{\hat{T}}_o &=& \hat{T}_1 - e_1 (\hat{T}_o - T_o) \nonumber
\end{eqnarray}
with $\hat{T}_o = \hat{T}$ and constant $e_o$ and $e_1$ leads to
$$
\ddot{\hat{\tau}} = (-e^{-\frac{E}{T}} - e_1) \dot{\hat{\tau}} - e_o \
\hat{\tau} + e_1 \dot{\tau}(t) + e_o \tau(t) 
$$ 
whose contraction behavior can now be tuned as in
\Ex{VanderPol}.}{chemical}
\Example{}{
Consider the system
$$
\dddot{x}\ = \ \sin (\dot{x})\ +\ 0.1 \sin t\ - \ 0.015
$$ 
with measurement $y=x$. Letting $e_0(y)=y$, $e_1(y)=4y$, and
$e_2(y)=3y$, the variational equation is 
$$
\delta \dddot{\hat{x}}+3\ \delta \ddot{\hat x}+(4-\cos(\dot{\hat x}))\
\delta \dot{\hat x}+\delta \hat{x}\ =\ 0
$$
The corresponding observer is illustrated in Figure
\ref{fig:example_4}.}{}

\subsection{Adding an indifferent system}

The analysis may be further simplified by using superposition to
compare the system to one whose contraction behavior is known
analytically. We illustrate this idea on second-order systems using
an indifferent added dynamics.  In principle, the approach can be
extended to higher-order systems as well as other types of added
dynamics.

Consider the indifferent system \cite{Lohm1}
$$
\delta \dot{\bf x} = i \Omega \delta {\bf x}
$$
with real and invertible $\Omega(\dot {\bf x}, {\bf x}, t)$. The above
corresponds to the second-order dynamics
$$
\delta \ddot{\bf x} = \dot{\Omega} \Omega^{-1} \delta \dot{\bf x} -
\Omega \Omega \delta {\bf x}
$$ 
which is thus itself indifferent.  We can write the reduced form of
Theorem \ref{th:higherordercontinuous} as
\begin{eqnarray} 
\frac{d}{dt}
\left(
\begin{array}{c}
  \delta \dot{\bf x} \\
  \delta     {\bf x}
\end{array} 
\right) &=&
\left(
\begin{array}{cc}
  - \Theta({\bf x}, \dot{\bf x}, t) & - \Omega \Omega(\dot {\bf x}, {\bf
    x}, t)  \\
  {\bf I} & {\bf 0}
\end{array} 
\right)
\left(
\begin{array}{c}
  \delta \dot{\bf x} \\
  \delta     {\bf x}
\end{array} 
\right) \nonumber \\ &=&
\left( \left(
\begin{array}{cc}
  \dot{\Omega} \Omega^{-1} & - \Omega \Omega \\
  {\bf I} & {\bf 0}
\end{array} 
\right)
+
\left(
\begin{array}{cc}
  - \dot{\Omega} \Omega^{-1} - \Theta & \ {\bf 0} \\
  {\bf 0} & \ {\bf 0}
\end{array} 
\right)
\right)
\left(
\begin{array}{c}
  \delta \dot{\bf x} \\
  \delta     {\bf x}
\end{array} 
\right) \nonumber
\end{eqnarray}
which thus corresponds to the superposition of an indifferent system with a
semi-contracting system of rate $-\Theta - \dot{\Omega} \Omega^{-1}$.

\begin{theorem}
The reduced form
\begin{equation} 
\delta \ddot{\bf x} + \Theta({\bf x}, \dot{\bf x}, t) \delta
\dot{\bf x} + \Omega \Omega(\dot {\bf x}, {\bf x}, t)\ \delta{\bf x} =
0 \nonumber
\end{equation}
is semi-contracting with rate $-\Theta - \dot{\Omega}
\Omega^{-1}$. The corresponding unreduced form (see Theorem
\ref{th:higherordercontinuous}) has an additional contraction rate
$-\frac{\bf F}{2}$.
\label{th:variationalenergy}
\end{theorem}

\Example{}{Consider the restricted Three-Body Problem \cite{ThreeBody}
$$
\ddot{\bf x} = 
2 \left(
\begin{array}{ccc}
  0  & 1 & 0 \\
  -1 & 0 & 0 \\
  0  & 0 & 0
\end{array}
\right)
\dot{\bf x} +
\left(
\begin{array}{ccc}
  0  & 1 & 0 \\
  1 & 0 & 0 \\
  0  & 0 & 0
\end{array}
\right)
{\bf x}
-\nabla V({\bf x})
$$ 
with state ${\bf x} = (x, y, z)^T$, potential energy $V({\bf x})
= -\nu \sqrt{y^2 + z^2 + (-1+x+y)^2}^{-1} - (1-\nu)
\sqrt{y^2+z^2+(x+y)^2}^{-1}$, $\nu$ the ratio of the smaller
mass to the larger mass, and the third body of zero mass.

Using ${\bf F} =  \left(
\begin{array}{ccc}
  0  & 1 & 0 \\
  -1 & 0 & 0 \\
  0  & 0 & 0
\end{array}
\right)$ in Theorem \ref{th:higherordercontinuous}, the reduced
variational dynamics is
$$
\delta \ddot{\bf x} = - \Delta V \delta {\bf x}
$$
Using Theorem \ref{th:variationalenergy}, the contraction behavior of
the three-body problem is the superposition of 

$\bullet$ $\ i \sqrt{\Delta V}$, which is indifferent for $\Delta V \ge
0$ and unstable otherwise.

$\bullet$ $\ -\frac{d \sqrt{\Delta V}}{dt} \sqrt{\Delta V}^{-1}$, where
a tightening (relaxing) potential force adds semi-contracting
(diverging)  behavior.
}{ThreeBodyProblem}

\section{Hamiltonian system dynamics} \label{Hamilton}

Consider the general $n$-dimensional Hamiltonian dynamics
\cite{Lovelock} with sum convention over the free index
$$
\ddot{x}^j + \gamma_{kh}^j \dot{x}^k \dot{x}^h = - H^{jh} f_h - D_h^j
\dot{x}^h
$$
with external forces $f_h(x^l, t)$, damping gain $D_h^j(x^l)$ and the
Christoffel term $\gamma^j_{kh} = \frac{1}{2} H^{mj} \left(
\frac{\partial H_{km}}{\partial x^h} + \frac{\partial H_{hm}}{\partial
x^k} - \frac{\partial H_{kh}}{\partial x^m} \right)$ (section 7.2. in
\cite{Lovelock}) of the symmetric inertia tensor $H_{lh}(x^m)$ and
where we use the convention $H^{mj} = H_{mj}^{-1}$. The variational
dynamics of the above is
$$
\delta \ddot{x}^j + \frac{\partial \gamma_{kh}^j}{\partial x^l} \delta
x^l \dot{x}^k \dot{x}^h + 2 \gamma_{kh}^j \dot{x}^k \delta \dot{x}^h =
- \frac{\partial (H^{jh} f_h)}{\partial x^l} \delta x^l - \frac{\partial
  D_h^j}{\partial x^l} \delta x^l \dot{x}^h - D_h^j \delta \dot{x}^h
$$
Let us now compute with the above the covariant time-derivative
(section 3.6. in \cite{Lovelock} and generalized contraction with metric
in \cite{Lohm1}) of the covariant velocity
variation $\delta \dot{x}^j + \gamma_{lh}^j \dot{x}^h \delta x^l$
with respect to the metric $H_{lh}$ as
\begin{eqnarray}
\frac{d}{dt} \left( \delta \dot{x}^j + \gamma_{lh}^j
\dot{x}^h \delta x^l \right) + \gamma_{ih}^j \dot{x}^h \left( \delta
\dot{x}^i + \gamma_{kl}^i \dot{x}^k \delta x^l \right) &=&
\nonumber \\ \delta \ddot{x}^j + \frac{\partial
\gamma_{lh}^j}{\partial x^k} \dot{x}^k \dot{x}^h \delta x^l +
\gamma_{lh}^j \ddot{x}^h \delta x^l + \gamma^j_{lh} \dot{x}^h \delta
\dot{x}^l + \gamma_{ih}^j \dot{x}^h \left( \delta \dot{x}^i +
\gamma_{kl}^i \dot{x}^k \delta x^l \right) &=& \nonumber \\
K^j_{lkh} \dot{x}^k \dot{x}^h \delta x^l - H^{jh} \left(
\frac{\partial f_h}{\partial x^l} - \gamma^k_{hl} f_k \right) \delta
x^l - D_h^j \left( \delta \dot{x}^h + \gamma^h_{kl} \dot{x}^k \delta
x^l \right)  \label{eq:varacceldynamics}
\end{eqnarray}
with curavture tensor $K^j_{lkh} = \frac{\partial
\gamma_{lh}^j}{\partial x^k} - \frac{\partial \gamma_{kh}^j}{\partial
x^l} + \gamma_{ih}^j \gamma_{kl}^i - \gamma_{li}^j \gamma_{kh}^i$
(section 7.3. in \cite{Lovelock}), covariant derivative of the
external forces $\frac{\partial f_h}{\partial x^l} - \gamma^k_{hl}
f_k$, and where we have assumed the covariant derivative of $D_h^j$,
that is $\frac{\partial D_h^j}{\partial x^l} + \gamma_{lk}^j D_h^k -
\gamma_{lh}^k D_k^j$ (see section 3.6 in \cite{Lovelock}), to vanish
which is usually the case for mechanical damping.

Note that according section 7.3 in \cite{Lovelock} corresponds
$K^i_{lkh} H_{ji} Y^k Y^h X^j X^l$ to the Riemannian or Gaussian
curvature of the 2-D subspace span by $X$ and $Y$. Hence the convexity
of $H$ orthogonal to $\dot{x}$ acts as a spring, whose gain
increases linearly with velocity. 

Also note that (\ref{eq:varacceldynamics}) can be used directly in
combination with Theorem 2 in \cite{Lohm5} to show under which
condition the covariant Hessian of the action $\phi$ becomes convex,
which implies contraction behavior.

However we use here Theorem \ref{th:higherordercontinuous} for ${\bf
F} = {\bf D}$, which is more general than the above (i.e. it allows to
assess complex solutions of the Hessian dynamics in Theorem 2 in
\cite{Lohm5}), to conclude:

\begin{theorem}
Consider for $t\ge0$ the $n$-dimensional ($n \ge 1$) dynamics  
with sum convention over the free index
\begin{equation}
\ddot{x}^j + \gamma_{kh}^j \dot{x}^k \dot{x}^h = - H^{jh} f_h - D_h^j
\dot{x}^h
\label{eq:Hamiltonian}
\end{equation}
with external forces $f^j(x^l, t)$, damping gain $D_h^j(x^l)$ with
covariant derivative $\frac{\partial D_h^j}{\partial x^l} +
\gamma_{lk}^j D_h^k - \gamma_{lh}^k D_k^j = 0$, and the Christoffel
term $\gamma^j_{kh} = \frac{1}{2} H^{mj} \left( \frac{\partial
H_{km}}{\partial x^h} + \frac{\partial H_{hm}}{\partial x^k} -
\frac{\partial H_{kh}}{\partial x^m} \right)$ of the symetric,
u.p.d. and invertible inertia tensor $H_{lh}(x^m)$.

Let us define a constant $\lambda \ge 0$ as the largest singular value
$\forall t \ge 0$ as
\begin{small}
\begin{equation} 
\left( -K_{ipkh} \dot{x}^k \dot{x}^h + \frac{\partial f_i}{\partial
x^p} - \gamma^h_{ip} f_h + H_{ih} \frac{D_k^h D_p^k}{4} \right) H^{ij}
\left( -K_{jlkh} \dot{x}^k \dot{x}^h + \frac{\partial f_j}{\partial
x^l} - \gamma^h_{jh} f_h + H_{jh} \frac{D_k^h D_l^k}{4} \right) \le
\lambda^2 H_{pl} \label{eq:characteristicHamiltonian}
\end{equation}
\end{small}
with curavture tensor $K_{jlkh} = H_{jo} \frac{\partial
\gamma_{lh}^o}{\partial x^k} - \frac{\partial \gamma_{kh}^o}{\partial
x^l} + \gamma_{ih}^o \gamma_{kl}^i - \gamma_{li}^o \gamma_{kh}^i$.

We can then conclude on contraction rate $- \frac{D_h^j}{2} +
\lambda$. \label{th:Hamiltoniancontinuous}
\end{theorem}

Also note that taking the double integral of the dynamics leads to an
exponential Lyaponov energy stability proof for the autonomous
case. In this sense Theorem 5 represents a variational extension of
the classical energy based Lyaponov proofs for autonomous systems.
Note that at a variational energy approach a tightening spring
(positive $\dot{\Omega}$) leads to semi-contraction behavior.

\Example{}{Let us now illustrate the simplicity of the stability
results using the inertia tensor as metric. Consider the Euler
dynamics of a rigid body, with Euler angles ${\bf x} = (\psi, \theta,
\phi)^T$ and measured rotation vector ${\bf \omega}$ in body
coordinates \cite{Goldstein}
\begin{equation}
\dot{\bf x} = \left(
\begin{array}{ccc}
  1 &          0 & - \sin \theta \\
  0 &  \cos \psi &   \cos \theta \sin \psi \\
  0 & -\sin \psi &   \cos \theta \cos \psi
\end{array}
\right)^{-1} {\bf \omega} \label{eq:rotationdynamics}
\end{equation}
The underlying energy is  
\begin{eqnarray}
h &=& \frac{1}{2} {\bf \omega}^T {\bf \omega} \nonumber \\
&=& \frac{1}{2}\dot{\bf x}^T 
\left(
\begin{array}{ccc}
  1             & 0 & -\sin \theta \\
              0 & 1 & 0 \\
  - \sin \theta & 0 & 1
\end{array}
\right)
\dot{\bf x} \nonumber
\end{eqnarray}
After a straightforward but tedious calculation we can compute
$$
\frac{d}{dt} \left( \delta {\bf x}^T {\bf H} \delta {\bf x} \right) = 0
$$
Thus, the Euler dynamics (\ref{eq:rotationdynamics}) is globally
indifferent. Note that this can also be seen from the
quaternion angular dynamics, whose Jacobian is
skew-symmetric~\cite{inertial}.}{Eulerdynamics}
 
\Example{}{The convexity of a the inertia tensor can act similarly to
a stabilizing potential force in the variational Hamiltonian dynamics
(\ref{eq:varacceldynamics}). Consider a rotating point mass of mass
$m$ on a ball with radius $R$. The Hamiltonian energy is
$$
h = \frac{m R^2}{2} \dot{\bf x}^T
\left(
\begin{array}{cc}
  1 &  0  \\
  0 &  \sin^2 \phi 
\end{array}
\right)
\dot{\bf x}
$$
with latitude $\phi$ and longitude $\psi$ in ${\bf x} = (\phi,
\psi)^T$. The curvature tensor can be computed e.g. with MAPLE as
$$
K_{jlkh} \dot{x}^k \dot{x}^h =
\frac{m R^2 \sin^2 \phi}{2} 
\left( \begin {array}{cc}  
-\dot{\psi}^2         & \dot{\psi} \dot{\phi} \\
\dot{\psi} \dot{\phi} & -\dot{\phi}^2  
\end {array} 
\right)
$$ which scales as the inertia tensor with $\frac{m R^2}{2}$ and is
negative orthogonal to the velocity and indifferent along the
velocity. Hence the convexitiy in Theorem
\ref{th:Hamiltoniancontinuous} acts under motion as a stabilizing
spring, that lets two moving neighboring trajectories oscillate around
each other.}{ball}

\Example{}{Theorem \ref{th:Hamiltoniancontinuous} can be used to
define observers or tracking controllers for time-varying Hamiltonian
systems. Consider a two-link robot manipulator, with kinetic energy
$$
\frac{1}{2}
\left( \begin{array}{cc} \dot{q}_1 & \dot{q}_2 \end{array} \right)
\left( \begin{array}{cc} 
a_1 + 2 a_2 \cos q_2 & a_2 \cos q_2 + a_3 \\
a_2 \cos q_2 + a_3 & a_3 
\end{array} \right)
\left( \begin{array}{c} \dot{q}_1 \\ \dot{q}_2 \end{array} \right)
$$
with $a_1 = m_1 l_{c1}^2 + I_1 + m_2(l_1^2+l_{c2}^2) + I_2$, $a_2 =
m_2 l_1 l_{c2}$ and $a_3 = m_2 L_{c2}^2 + I_2$ and $-\pi \le q_1, q_2
\le \pi$.  Let us assume that $q^j$ is measured and define the
observer
\begin{eqnarray}
\dot{\hat{\omega}}^j + \gamma_{kh}^j(\hat{q}^j) \dot{\hat{q}}^k
\dot{\hat{q}}^h &=& - H^{hj} f_h \nonumber \\ \dot{\hat{q}}^j &=&
\hat{\omega}^j + d_h^j ( q^h - \hat{q}^h ) \nonumber
\end{eqnarray}
with the external forces
$$
f_h = 
\left( 
\begin{array}{c}
g (m_1 l_{c1} + m_e l_1) \cos q_1 + g m_e l_{ce} \cos(q_1 + q_2) +
\tau_1 \\ g m_e l_{ce} \cos(q_1 + q_2) + \tau_2
\end{array}
\right)
+ k_h^i \left(\hat{q}^i -q^i \right)
$$  
and external torques $\tau_1$, $\tau_2$. The above is equivalent to
(\ref{eq:Hamiltonian}) 
$$
\ddot{\hat{q}}^j + \gamma_{kh}^j(\hat{q}^j) \dot{\hat{q}}^k
\dot{\hat{q}}^h = - H^{jh} \left( f_h - d_{hk} \dot{q}^k \right)
- d_h^j \dot{\hat{q}}^h  
$$
where the covariant derivative of a constant scalar $d_h^j$
vanishes. The curvature tensor can be computed e.g. with MAPLE as
$$
K_{jlkh} \dot{x}^k \dot{x}^h =
\frac{ \left( a_1 a_3 - a_3^2 - a_2^2 \right) a_2 \cos q_2} {a_1 a_3
  -a_2^2 \cos^2 q_2 -a_3^2}
\left( \begin {array}{cc}  
-\dot{q}_2^2 & \dot{q}_1 \dot{q}_2 \\
\dot{q}_1 \dot{q}_2 & -\dot{q}_1^2  
\end {array} 
\right)
$$
The curvature is for $a_1 a_3 \ge a_2^2 + a_3^2$ convex (concave) for
$-\frac{\pi}{2} \le q_2 \le \frac{\pi}{2}$ ($-\frac{\pi}{2} \ge q_2 \
or \ q_2 \ge \frac{\pi}{2}$) and accordingly (de)-stabilizes the
dynamics when the arm is retracted (extended). Let us now compute the
covariant derivative of the external forces
$$
\frac{\partial f_i}{\partial x^p} + \gamma^h_{ip} \left( f_h - d_{hk}
\dot{q}^k \right) = -k_p^i + \gamma^h_{ip} \left( f_h - d_{hk}
\dot{q}^k \right)
$$
with $-\gamma^1_{11} = \gamma^2_{12} = \gamma^2_{21} = \gamma^2_{22} =
\frac{ \left( a_2 \cos q_2 + a_3 \right) a_2 \sin q_2} {a_1 a_3 -a_2^2
\cos^2 q_2 -a_3^2}$, $\gamma^1_{12} = \gamma^1_{21} = \gamma^1_{22} =
- \frac{a_3 a_2 \sin q_2} {a_1 a_3 -a_2^2 \cos^2 q_2 -a_3^2}$ and
$\gamma^2_{11} = \frac{ \left( a_1 + 2 a_2 \cos q_2 \right) a_2 \sin
q_2} {a_1 a_3 -a_2^2 \cos^2 q_2 -a_3^2}$.

The spring gain $k_p^i$ stabilizes the system, whereas the supporting
force $f_h - d_{hk} \dot{q}^k$ can (de)-stabilize the system
proportional to the magnitude of the supporting force. Note that $f_h
- d_{hk} \dot{q}^k$ can (de)-stabilize a curved system is unavoidable
since here no constant or parallel (force) vectors exist, whose
covariant derivative vanishes \cite{Lovelock}.

Computing $\lambda$ from (\ref{eq:characteristicHamiltonian}) then
allows with Theoreom \ref{th:Hamiltoniancontinuous} to compute bounds
on the velocity $\dot{q}^j$ and external forces $f_h - d_{hk}
\dot{q}^k$ for which global contraction behavior can be concluded.

System responses to a control input $\tau_i = (\sin t, \cos 5 t)$,
initial conditions $q^i(0) = (-\frac{\pi}{2}, \pi)$ rad, $\dot{q}^i(0)
= (3, -3)$ rad/s, $\hat{q}^i(0) = (- \frac{\pi}{2}, \pi)$ rad,
$\hat{\dot{q}}^i(0) = (-5, 5)$ rad/s and parameters $m_1 = 1$ kg, $l_1
= 1$ m, $m_e = 2$ kg, $I_1 = 0.12$ kgm$^2$, $l_{c1} = 0.5$ m, $I_e =
0.25$ kgm$^2$, $l_{ce} = 0.6$ m, $d_h^j = 5$ Nms/rad, $k_h^i = 5$
Nm/rad are illustrated in figure \ref{fig:robotqhat1} and
\ref{fig:robotwhat1}. The solid lines represent the real plant, and
the dashed lines the observer estimate.

\begin{figure}
\begin{center}
\includegraphics[scale=0.3]{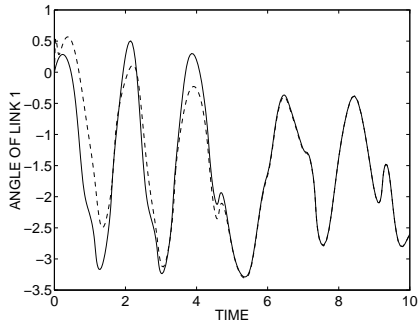}
\end{center}
\caption{Positions of two-link robot}
\label{fig:robotqhat1}
\end{figure}

\begin{figure}
\begin{center}
\includegraphics[scale=0.3]{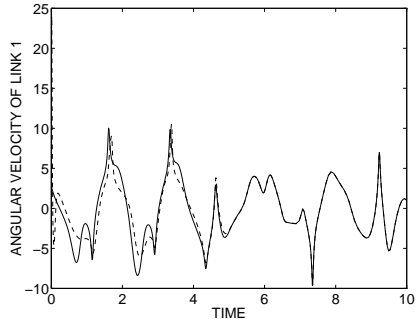}
\end{center}
\caption{Velocities of two-link robot}
\label{fig:robotwhat1}
\end{figure}
The above observers provide a simple alternative to current design
methods [see e.g., Berghuis and Nijmeyer, 1993; Marino and Tomei,
1995], and guarantees local (for bounded velocities and time-varing
inputs) exponential convergence.

Note that Theoreom \ref{th:Hamiltoniancontinuous} can also be used to
bound the diverging behavior, caused by the concave inertia when the
robot arm is pointing inwards, of the double inverted pendulum, when
no damping or stabilizing potential force is applied.}{robot} 

\Example{}{Biology found a solution to the problem that a time-varying
supporting torque can destabilize a system. 

Recently, there has been considerable interest in analyzing feedback
controllers for biological motor control systems as combinations of
simpler elements, or motion primitives.  For instance \cite{Bizzi} and
\cite{Mussa} stimulate a small number of areas (A, B, C, and D) in a
frog's spinal cord and measure the resulting torque angle relations.

Force fields seem to add when different areas are stimulated at the
same time so that \cite{Bizzi} and \cite{Mussa} propose the following
biological control inputs
$$
  f_i = - \sum_{l=1}^n k_l(t) f_{il} (q^j)
$$
where each single torque $k_l(t) f_{il} (q^j)$ results from the
stimulation of area $l$ in the spinal cord. With force measurements
the above authors did show that the covariant derivative
$\frac{\partial f_{il}}{\partial x^p} + \gamma^h_{ip} f_{hl}$ of
$f_{il}$ with respect to the inertia tensor of the frog's leg or body
is uniformly positive definite.

Likely candidates for $k_i(t)$ are positive upper and lower bounded
sigmoids and pulses and periodic activation patterns.

Using Theorem \ref{th:Hamiltoniancontinuous} or the discussion in
\Ex{robot} with sufficient damping then allows to compute a maximal
$\dot{q}^j$ for which exponential convergence to a single motion is
guaranteed.

Note that the achievement of tracking control with a proportional gain
$k_l(t)$ rather than an additional supporting force as in \Ex{robot}
has the advantage that the supporting force has no impact on the
contraction behavior anymore.}{motionprimitives.}

\section{Concluding remarks}

The research in this paper can be extended in several directions, as
the development suggests.

Some of the extensions will likely require the combination of the
above results with a simplifying metric pre-transformation, as
mentioned in section 3.1.  In particular, classical transformation
ideas in nonlinear control such as feedback linearization and
flatness~\cite{flatness} typically use linear time-invariant target
dynamics, while the framework provided in this paper should allow
considerably more flexibility. This, combined with the fact that a
metric transformation such as $\ \delta {\bf z} = {\bf \Theta} \delta
{\bf x}\ $ need not be integrable (i.e. does not require an explicit
${\bf z}$ to exist), could potentially lead to useful generalisations
of these methods.

{\bf Acknowledgement}\ \ The authors are grateful to Yong Zhao for
performing the simulations and for stimulating discussions, and to Wei
Wang for thoughtful comments and suggestions.

\newpage

\end{document}